\documentclass[12pt]{article}
\usepackage[dvips]{graphicx}
\usepackage{pdproc}

  \makeatletter 
  \def\@cite#1{[#1]} 
  \makeatother    
\begin{document}
\pagestyle{plain}

\title{Supersymmetric Dark Matter 2004\footnote{Plenary talk at {\it SUSY04},
  Tsukuba, Japan, June 2004}}

\author{MANUEL DREES}

\address{ 
Physikalisches Institut, Universit\"at Bonn, \\
Nussallee 12, 53115 Bonn, Germany }

\abstract{ Recent cosmological data allow to determine the universal Dark
  Matter (DM) density to a precision of about 10\%, {\em if} a simple,
  well--motivated ansatz for the spectrum of primordial density perturbations
  is correct. Not surprisingly, a thermal neutralino $\tilde \chi_1^0$ will
  have the correct relic density only in ``small'' regions of parameter space.
  In particular, for fixed values of the other parameters, the allowed region
  in the $(m_0, m_{1/2}$ plane (in mSUGRA or similar models) seems quite
  small, if standard assumptions about the Universe at temperature $T \simeq
  m_{\tilde \chi_1^0}/10$ are correct. I argue that the allowed parameter
  space is actually still quite large, when all uncertainties are properly
  taken into account. In particular, the current lower limits on sparticle and
  Higgs masses that can be derived within mSUGRA do {\em not} change
  appreciably when the DM relic density constraint is imposed. I also show
  that deviating from mSUGRA does not alleviate the finetuning required to
  obtain the correct relic density, unless one also postulates a non--standard
  cosmology. Finally, I briefly discuss claimed positive evidence for
  particle Dark Matter.}

\normalsize\baselineskip=15pt

\section{Introduction}

We live in an era of precision cosmology: the statistical errors of several
cosmological measurements have now reached the percent level. A prime example
is the celebrated ``WMAP'' data (which include data from smaller experiments
as well) on the cosmic microwave background (CMB). Using these data in
combination with other observations of the large--scale structure of the
Universe, {\em and} making reasonable assumptions about the early Universe,
one can extract several quantities that are of interest to (astro)particle
physicists \cite{md1}. These include in particular the scaled non--baryonic
Dark Matter (DM) density $\Omega_{\rm DM}$ multiplied with the scaled Hubble
constant $h$:
\begin{equation} \label{om1}
\Omega_{\rm DM} h^2 = 0.113 \pm 0.009\,.
\end{equation}

As emphasized above, one has to make assumptions about the early Universe in
order to derive Eq.(\ref{om1}). In particular, one has to assume a functional
form for the ``power spectrum'' (essentially, the Fourier spectrum) of
primordial density fluctuations which seed structure formation.  The result
(\ref{om1}) is valid for a nearly scale--invariant power spectrum, with slowly
varying spectral index \cite{md1}.  This ansatz is well motivated from
inflation \cite{md2}.  Moreover, a slightly different ansatz (a simple power
law) gives very similar results.  However, recall that the CMB spectrum
essentially measures a function of one variable, usually taken to be the
angular mode variable $\ell$.  If the primordial power spectrum, another
function of one variable, is left unconstrained, a measurement of the CMB
anisotropies could {\em not} determine {\em any} parameters other than the
initial power spectrum, which is of little immediate interest to particle
physicists. 

Of course, the fact that a simple, well--motivated ansatz for the primordial
density fluctuations leads to a reasonable description of the CMB (and
related) data for some values of the relevant cosmological parameters
(including $\Omega_{\rm DM}$) is highly nontrivial. However, the description
of these data, including Eq.(\ref{om1}), also has a couple of puzzling
features. In particular, on very large angular scales the anisotropies are
somewhat smaller than expected. This $\sim 2 \sigma$ discrepancy has triggered
numerous speculations, ranging from the quite prosaic (e.g. models with more
than  one inflaton field \cite{md3}) to the fairly exotic (e.g. models where
the Universe has nontrivial topology \cite{md4}).

Secondly, the same fit that produces Eq.(\ref{om1}) also requires that the
Universe should have been re--ionized at redshift $z = 17 \pm 5$ \cite{md1}.
The time when nuclei and electrons first combined to firm a neutral gas
defines the famous surface of last scattering of the CMB photons (at $z \sim
10^3$), since a neutral gas is essentially transparent while an ionized plasma
is not. In today's Universe most gas is again not ionized. The standard
explanation for ionization is that the earliest generation of stars, called
``population III'' by astronomers\footnote{Astronomers count backwards. The
  most recent generation of stars, including our Sun, is called population I.
  Old population II stars can be found e.g. in globular clusters. No
  population III stars have been identified, but they have to exist (or at
  lest, to have existed), since the initial chemical composition of population
  II stars differs significantly from that of the primordial post--BBN
  Universe.}, emitted enough UV radiation to re--ionize the Universe. However,
in the quite recent past the existence of stars (in galaxies) at redshift $z
\geq 5$ was thought to be quite challenging for standard CDM cosmology; now a
significant density of stars at redshift $z \simeq 17$ seems required. This
has led to speculations \cite{md5} that the early re--ionization might have
been due to the (radiative) decay of some long--lived massive particles. Note
that these particles would have contributed to $\Omega_{\rm DM}$ at the time
of last scattering of CMB photons, but would have decayed by now.

The upshot of this lengthy discussion is that Eq.(\ref{om1}) should be taken
with a grain of salt. Note that the error given there is purely statistical.
It should be clear that this measurement also has a systematic uncertainty,
but I do not know how to estimate it. For the remainder of this write--up I
will assume that $\Omega_{\rm DM}$ falls in the 99\% c.l. confidence range
\begin{equation} \label{om2}
0.087 \leq \Omega_{\rm DM} h^2 \leq 0.138.
\end{equation}

A particle $\chi$ has to satisfy several fairly obvious conditions in order to
qualify as a DM candidate: it must be very long--lived, $\tau_\chi > 10^{10}$
yrs; it must be electrically and (most likely) color neutral, since otherwise
it would bind to nuclei, forming exotic isotopes in conflict with experimental
limits \cite{md6}; its scattering cross section on nucleons must be below the
experimental limits from direct WIMP searches \cite{md7}; and its relic
density must fall in the range (\ref{om2}). If we want $\chi$ to be a
sparticle in the visible sector of the MSSM, the first three requirements
uniquely single out the lightest neutralino, $\chi = \tilde\chi_1^0$; a
sneutrino would violate the direct WIMP search limits \cite{md7} by several
orders of magnitude \cite{sneut}. However, $\chi$ could also reside in the
``hidden sector'' thought to be responsible for the spontaneous breaking of
supersymmetry. In the following section I will discuss the simplest scenario,
thermal $\tilde \chi_1^0$ DM in constrained models assuming standard
cosmology, while sec.~3 deals with other possibilities. Some claimed positive
observational evidence for WIMP DM will be discussed in Sec.~4, before
concluding in Sec.~5.

\section{The simplest scenario}

The lightest neutralino $\tilde \chi_1^0$ owes its popularity as DM candidate
to several features. Under some (rather mild) assumptions about the early
Universe the $\tilde \chi_1^0$ relic density can be calculated as function of
particle physics parameters only; this gives the desired value for some
regions of parameter space even in constrained models like mSUGRA ($\equiv$
CMSSM) \cite{md8}. Similar statements also hold for other thermal WIMP
candidates, but they have little \cite{lkp} or no \cite{scalar} independent
motivation from particle physics.\footnote{Note in particular that models with
  ``universal'' extra dimensions, which contain a DM candidate, do not even
  pretend to solve the hierarchy problem.} Moreover, the hypothesis that
$\tilde \chi_1^0$ forms the DM in our galaxy can be tested experimentally; in
fact, as of this writing there are at least two claims for a positive signal
(see below).

In the post--WMAP era the neutralino DM candidate has nevertheless come to be
viewed more critically. This can at least partly be explained sociologically:
all simple calculations involving $\tilde \chi_1^0$ DM have been performed,
and phenomenologists (like me) have to keep occupied. One complaint is that
the size of the allowed parameter space shrinks drastically when the
constraint (\ref{om2}) is imposed, see Fig.~1. This can be dismissed out of
hand: it is not surprising, and is in fact highly desirable, that a precision
measurement reduces the size of the viable parameter space. In the limit of
vanishing experimental error a measurement should reduce the dimension of the
parameter space by one. (For example, the measurement of $M_Z$ is generally
used to determine $|\mu|$ in mSUGRA.)

\begin{figure}[htb]
\begin{center}
\vspace*{-1cm}
\rotatebox{270}{\includegraphics*[width=8.cm]{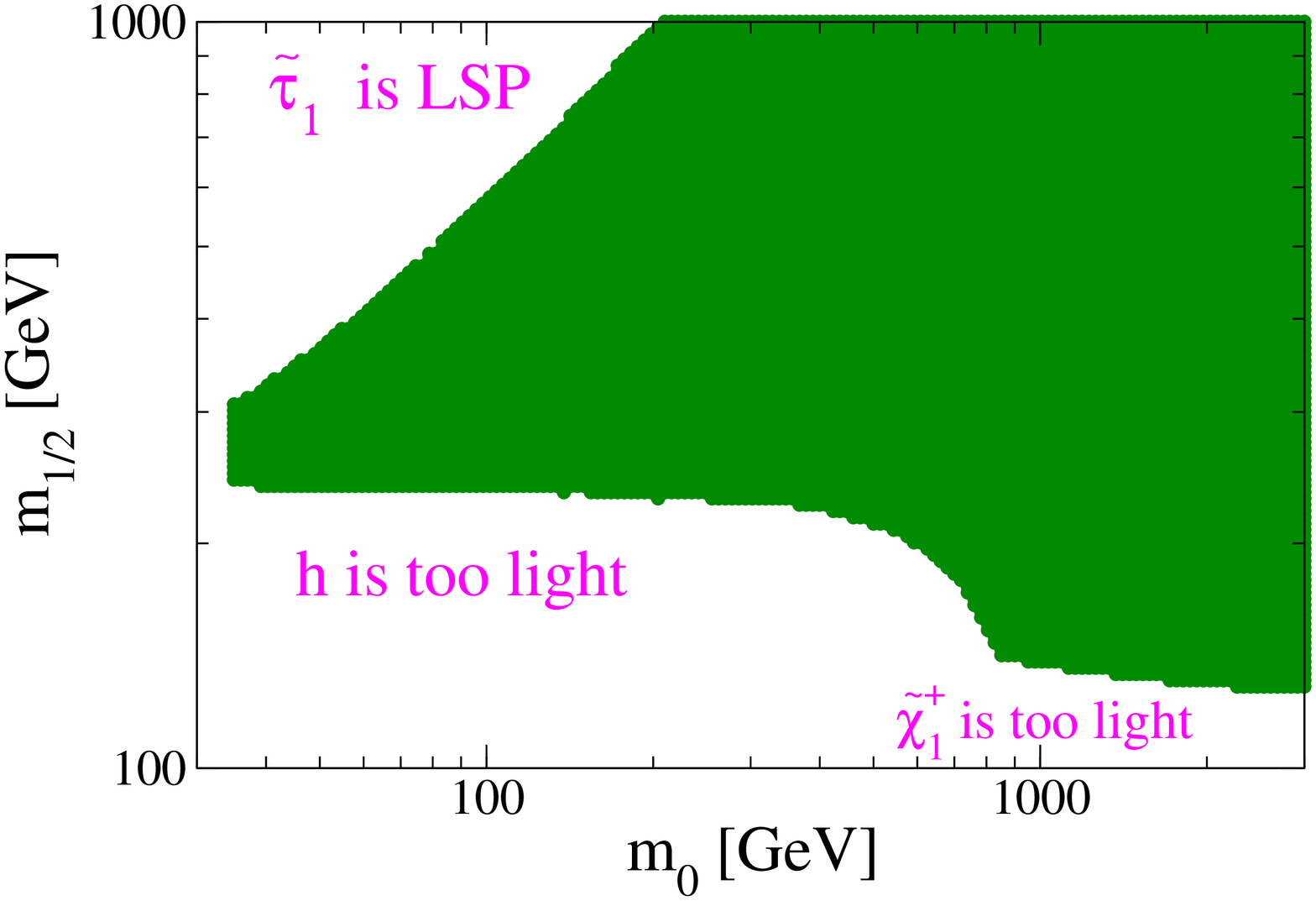}} \\
\vspace*{-.7cm}
\rotatebox{270}{\includegraphics*[width=8.cm]{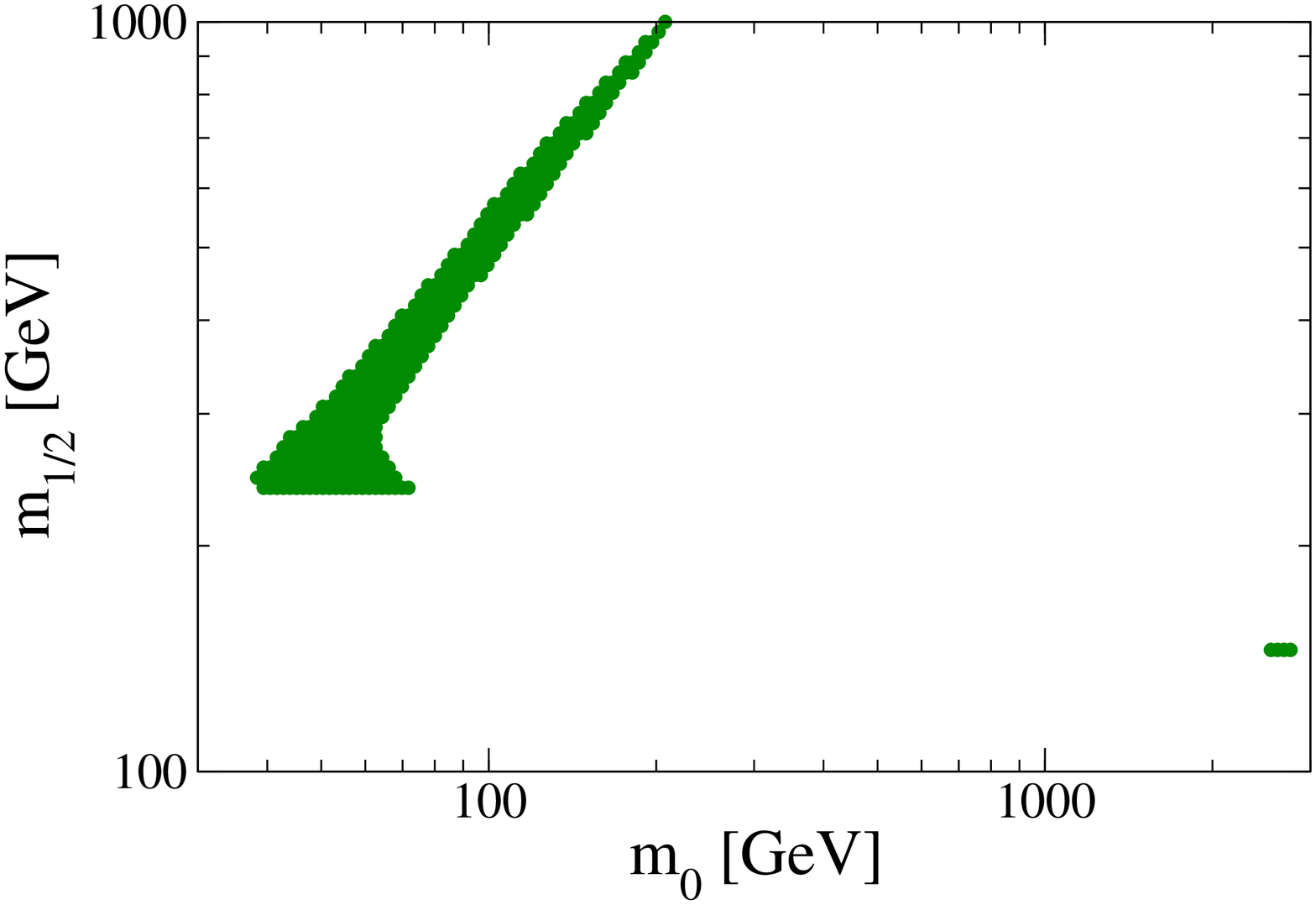}} \\
\caption{%
The allowed region in the $(m_0, m_{1/2})$ plane in mSUGRA with $A_0 = 0,\,
\tan\beta = 10, \, m_t = 178$ GeV and $\mu > 0$ without (top) and with
(bottom) the DM constraint of Eq.(2).}
\label{fig1}
\end{center}
\end{figure}

However, for the time being we are still quite far away from a situation with
negligible experimental errors. In fact, both the experimental errors on input
quantities (like the top mass $m_t$) and experimental measurements (like the
constraint (\ref{om2}) and the magnetic moment of the muon \cite{gmu}), and
theoretical uncertainties of the spectrum calculation should be taken
seriously when assessing the currently allowed parameter space of mSUGRA (or
any other model).

This is illustrated by Fig.~2, which shows the lower bounds on a few sparticle
and Higgs masses in mSUGRA for $\tan\beta = 20$. For the first set of points I
have set $m_t = 178$ GeV, $A_0 = 0$, $\mu > 0$, and scanned over $m_0$ and
$m_{1/2}$\footnote{I follow the notation of \cite{ddk1}.}, discarding all
points that violate any of the sparticle or Higgs production limits from LEP
\cite{md6}. The spectrum has been calculated with the latest version of
Suspect \cite{md9}. Note that this first set of points does not include the
constraint (\ref{om2}), nor does it include any constraints on $Br(b
\rightarrow s \gamma)$ (since this latter constraint can be evaded \cite{md10}
by introducing some $\tilde s -\tilde b$ mixing, without significantly
changing anything else in the spectrum). I did impose a mild version of the
$g_\mu$ constraint \cite{gmu}, obtained by taking the envelope of the $2
\sigma$ regions calculated using $\tau$ decay and $e^+e^-$ annihilation data,
respectively, when evaluating the SM prediction for $g_\mu$. This gives
\begin{equation} \label{gmu}
-5.7 \cdot 10^{-10} \leq a_\mu^{\rm SUSY} \leq 47.1 \cdot 10^{-10}\, .
\end{equation}

\begin{figure}[htb]
\begin{center}
\rotatebox{270}{\includegraphics*[width=13cm]{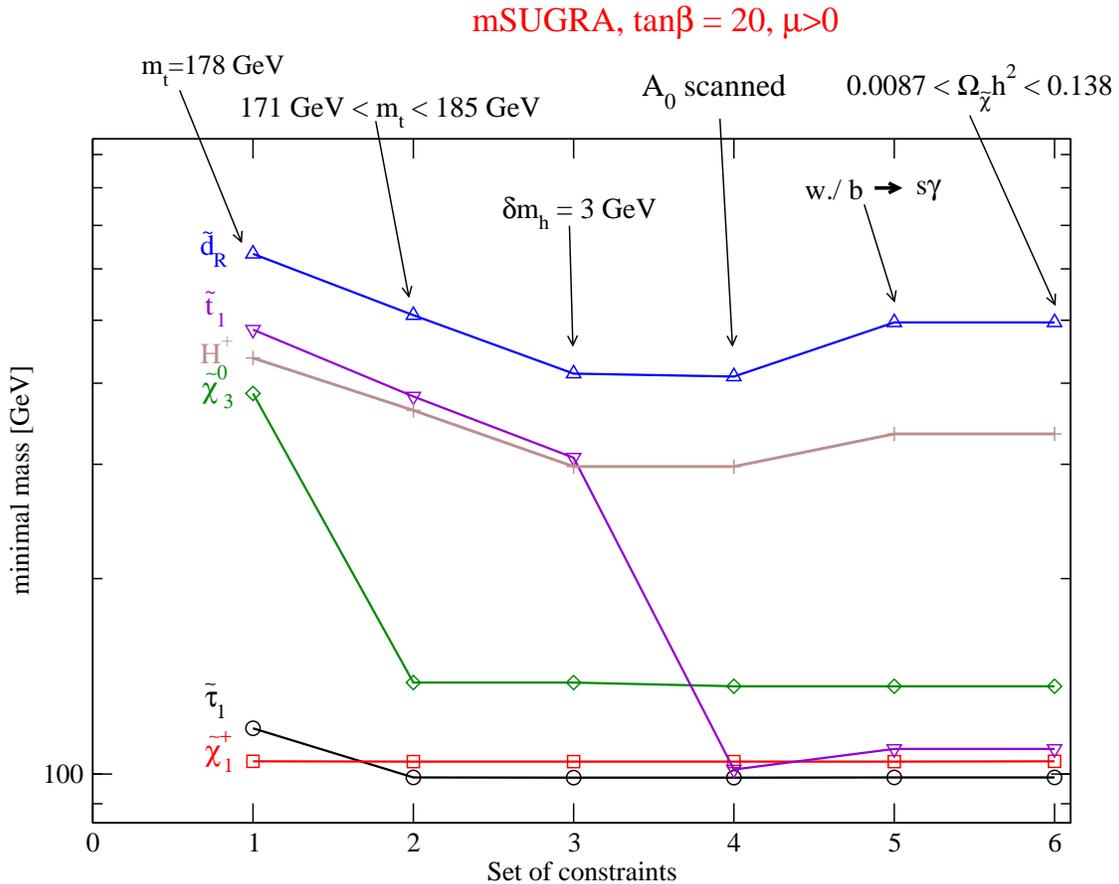}}
\caption{%
Lower bounds on some sparticle and Higgs boson masses predicted by mSUGRA
under various assumptions; see text for details.}
\label{fig2}
\end{center}
\end{figure}

We see that under these assumptions mSUGRA requires most sparticles to have
masses well above the direct experimental lower bounds. Only the chargino mass
saturates the LEP value of about 104.5 GeV. For example, first generation
squark masses are required to lie above 630 GeV. Recall, however, that I have
more or less arbitrarily fixed various parameters when determining these lower
bounds. In particular, the top mass comes with an error (from the direct
measurement) of 4.3 GeV, leading to a 90\% c.l. allowed range
\begin{equation} \label{mt}
171 \ {\rm GeV} \leq m_t \leq 185 \ {\rm GeV}.
\end{equation}
Allowing $m_t$ to lie anywhere in this range significantly reduces the lower
bounds. This is true in particular for the heavier neutralinos and chargino:
the bound on $m_{\tilde \chi_3^0}$ drops from about 380 to 140 GeV; the latter
is the lower bound that holds in a general MSSM with gaugino mass unification,
given the constraint $m_{\tilde \chi_1^\pm} > 104.5$ GeV. This big jump occurs
since for smaller $m_t$ the so--called focus point of hyperbolical branch
region \cite{md12} becomes accessible again, which combines large $m_0$ with
rather small $\mu$. Note that now also the $\tilde \tau_1$ mass is directly
given by its LEP lower bound of about 98 GeV (assuming an upper bound on the
$\tilde \tau_1$ pair production cross section; presumably somewhat smaller
$m_{\tilde \tau_1}$ are allowed for small $\tilde \tau_1 - \tilde \chi_1^0$
mass splitting). The lower bounds on the masses of the other scalars are also
reduced, chiefly because larger $m_t$ mean larger values of the mass of the
light Higgs boson $h$, allowing to reduce the sparticle mass scale
\cite{md11}.

In the next step I have introduced a theoretical uncertainty of 3 GeV on the
calculation of $m_h$ \cite{md13}. In practice this moves the LEP Higgs limit
down to about 111 GeV (unless $\tan\beta$ is very large; see below), which
again reduces the lower bounds on scalar masses.

So far I have kept $A_0 = 0$ fixed. Scanning over this parameter, subject to
the requirement that the weak--scale (!) scalar potential should not have
deeper minima breaking color or charge.\footnote{In other words, the
  so--called ``UFB'' constraints \cite{md14} are not imposed here. They
  significantly reduce the parameter space even for $A_0=0$, but even if a
  ``UFB'' minimum exists, our (false) vacuum is usually extremely long--lived.}
This leads to a further mild reduction of the SUSY mass scale needed to
satisfy the Higgs search limits. Much more significantly, it allows large
$\tilde t_L - \tilde t_R$ mixing, thereby moving the bound on $m_{\tilde t_1}$
down to its lower limit from LEP. (Tevatron limits do not apply, since the
mass splitting with the LSP is not large enough.) 

For the fifth set of bounds I have in addition imposed the constraint
\cite{md15} 
\begin{equation} \label{bsg}
2.65 \cdot 10^{-4} \leq Br(b \rightarrow s \gamma) \leq 4.45 \cdot 10^{-4}\,.
\end{equation}
For the given moderately large value of $\tan\beta$ this has some impact on
the lower bounds of the heavy scalars, e.g. increasing the limit on the mass
of first generation squarks by about 20\% to just below 500 GeV. On the other
hand, now finally also including the DM relic density constraint (\ref{om2})
has almost {\em no} impact on the lower bounds shown in Fig.~1. Table~1 shows
that this is also essentially true for the absolute lower bounds one derives
in mSUGRA after also scanning over $\tan\beta$. Introducing the constraint
(\ref{om2}) increases the lower bound on the LSP mass by some 5 GeV (to a value
close to $m_{h,\rm min}/2$, to benefit from enhanced $\tilde \chi_1^0$
annihilation through $h$ exchange), but none of the other lower bounds moves
significantly.  These mSUGRA lower bounds are quite close to those one would
obtain within a more general MSSM, as long as one keeps the gaugino masses
unified at the GUT scale, and requires that all squared squark and slepton
masses are non--negative up to that scale.

\begin{table}[h]
\begin{center}
\caption{
  Absolute lower bounds on some sparticle and Higgs masses (in GeV) in mSUGRA
  after scanning over the entire allowed parameter space, without and with
  the DM constraint (2); see text for further details.}
\vspace*{3mm}
\begin{tabular}{|c||c|c|}
\hline
Particle & W/o DM constraint& With DM constraint\\
\hline \hline
$\tilde \chi_1^0$ & 50.7 & 55.8 \\
$\tilde \chi_1^\pm$ & 104.5 & 104.5 \\
$\tilde \chi_3^0$ & 135.1 & 136.5 \\
\hline
$\tilde \tau_1$ & 98.7 & 98.7 \\
$h$ & 91.0 & 91.0 \\
$H^\pm$ & 128.4 & 128.4 \\
\hline
$\tilde g$ & 371 & 384 \\
$\tilde d_R$ & 411 & 411 \\
$\tilde t_1$ & 102 & 102 \\
\hline
\end{tabular}
\label{table1} 
\end{center}
\end{table}

I should emphasize that these bounds are saturated in quite different regions
of pa\-ra\-me\-ter space. As noted earlier, the masses of the heavier $\tilde
\chi$ states are minimized for $m_0 > 1$ TeV and keeping $m_t$ close to the
lower end of the range (\ref{mt}), while the first generation squark and
slepton masses are minimal for the largest allowed $m_t$, small $m_0$ and
$m_{1/2}$, and moderate $\tan\beta$ (where the $g_\mu$ and $b \rightarrow s
\gamma$ constraints are not significant). Finally, the lower bounds on the
Higgs masses are saturated at $\tan\beta = 60$; no solution is found for
significantly larger values of this parameter.

We thus see that the additional restrictions of parameter space from the
constraint (\ref{om2}) applied to a stable $\tilde \chi_1^0$ as DM in some
sense are not very severe; at least they do not change significantly the lower
bound on the mass of any new particle predicted by mSUGRA. Nevertheless it is
often argued that this constraint leads to additional finetuning, since it can
be satisfied only in ``peculiar'' regions of parameter space: the ``focus
point'' region; the co--annihilation region, with small mass splitting between
the LSP and either the $\tilde \tau_1$ \cite{md17} or $\tilde t_1$
\cite{md18}; or the ``Higgs pole'' or ``funnel'' region, where $2 m_{\tilde
\chi_1^0} \simeq m_A$ \cite{md19}. The first two of these regions are very
close to the edge of theoretically forbidden regions (by the requirement of
consistent electroweak symmetry breaking and of a neutral LSP, respectively).
Moreover, in all these cases $\Omega_{\rm DM} h^2$ depends very sensitively
on some input parameter(s): on $m_t$ and $m_{1/2}$ in the focus point
region; on the LSP--sfermion mass splitting in the co--annihilation regions;
and on $2 m_{\tilde \chi_1^0} - m_A$ in the Higgs pole region. 
  
This objection to $\tilde \chi_1^0$ as DM triggered a fair amount of work in
recent years on non--minimal scenarios, some of which I will briefly discuss
in the next section. However, I first want to point out that the ``bulk''
region, where a bino--like $\tilde \chi_1^0$ has sufficiently large
annihilation cross section due to the exchange of sufficiently light
sfermions (mostly sleptons) without unduly strong dependence on input
parameters, still exists if one takes the uncertainties discussed above
seriously. An example is given in Table~3, which saturates the limits on
$m_t, \, m_h$ and $m_{\tilde \tau_1}$, but gets $\Omega_{\rm DM}$ ``right on
the money'' with a sparticle spectrum in easy range of near--future
experiments.

\begin{table}[h]
\begin{center}
\caption{
  Example for an allowed mSUGRA parameter space point in the ``bulk'' region.
  All mass parameters are in GeV. The first three rows give input parameters,
  the remaining quantities are calculated.}
\vspace*{3mm}
\begin{tabular}{|c|c||c|c|}
\hline
Quantity & Value & Quantity & Value \\
\hline \hline
$m_t$ & 185 & $m_0$ & 70.3 \\
sign($\mu$) & +1 & $m_{1/2}$ & 181.3 \\
$\tan\beta$ & 6 & $A_0$ & -375 \\
\hline
$m_{\tilde \chi_1^0}$ & 67.9 & $m_{\tilde e_R}$ & 108 \\
$m_{\tilde \chi_2^0}$ & 127 & $m_{\tilde e_L}$ & 152 \\
$m_{\tilde \chi_3^0}$ & 332 & $m_{\tilde \nu_e}$ & 130 \\
$m_{\tilde \chi_4^0}$ & 351 & $m_{\tilde \tau_1}$ & 99 \\
$m_{\tilde \chi_1^\pm}$ & 126 & $m_{\tilde \tau_2}$ & 155 \\
$m_{\tilde \chi_1^\pm}$ & 351 & $m_{\tilde \nu_\tau}$ & 129 \\
\hline
$m_h$ & 111& $m_A$ & 355 \\
$m_H$ & 356& $m_{H^\pm}$ & 128.4 \\
\hline
$m_{\tilde g}$ & 452 & $m_{\tilde t_1}$ & 240 \\
$m_{\tilde d_R}$ & 410 & $m_{\tilde t_2}$ & 470 \\
$m_{\tilde u_R}$ & 411 & $m_{\tilde b_1}$ & 377 \\
$m_{\tilde d_L}$ & 431 & $m_{\tilde b_2}$ & 412 \\
\hline
$\Omega_{\tilde \chi_1^0} h^2$ & 0.114 & $Br(b \rightarrow s \gamma)$ & $2.7
\cdot 10^{-4}$ \\
\hline
\end{tabular}
\label{table2} 
\end{center}
\end{table}

\section{Nonminimal scenarios}

We saw in Fig.~1 that in mSUGRA most of the $(m_0, m_{1/2})$ plane is excluded
by the DM constraint (\ref{om2}) if the other parameters are held fixed. The
reason is that ``generically'', $\tilde \chi_1^0$ is bino--like in mSUGRA.
This means that it mostly annihilates from a $P-$wave initial state (which
increases the DM density by a factor $\sim 7$, relative to annihilation from
an $S-$wave); only annihilates via $U(1)_Y$ interactions, which have the
smallest gauge coupling of the three SM factor groups; and mostly annihilates
through the exchange of sfermions in the $t-$ and $u-$channel, which are often
significantly heavier than $\tilde \chi_1^0$.  Evidently the annihilation
cross section could be increased, and hence the relic density reduced, if some
or all of these suppression factors could be removed, without having to change
the cosmology.

In particular, $\tilde \chi_1^0$ can be made higgsino--like in a number of
ways: by allowing the Higgs soft masses at the input (GUT) scale to exceed the
sfermion masses \cite{md20}; by reducing the input scale from $M_X \simeq 2
\cdot 10^{16}$ GeV to some value around $10^{10}$ GeV \cite{md21}; or by
reducing the gluino mass at the input scale relative to the electroweak
gaugino masses \cite{md22}. Similarly, $\tilde \chi_1^0$ will be wino--like if
the $SU(2)$ gaugino mass $M_2$ is smaller than the $U(1)_Y$ gaugino mass $M_1$
at the weak scale, which requires $M_2 < M_1/2$ at scale $M_X$
\cite{md23}. Finally, the biggest (effective) DM annihilation cross section
arises if $\tilde \chi_1^0$ is nearly degenerate with a strongly interacting
sparticle, e.g. $\tilde t_1$, leading to strong co--annihilation \cite{md18};
this option even exists in mSUGRA. The trouble with all these modifications is
that they do not really improve the situation. In particular, they do not
increase the fraction of the (now enlarged) parameter space where the
constraint (\ref{om2}) is satisfied; nor do they reduce the finetuning
required to satisfy this constraint.

This is illustrated in Fig.~3, which shows $\Omega_{\tilde \chi_1^0} h^2$ as a
function of one parameter. This parameter is $A_0$ (which controls $m_{\tilde
  t_1}$) for the (red) dot--dashed curve; the ratio of squared Higgs and
sfermion soft masses for the dashed (blue) curve; and the ratio of $U(1)_Y$
and $SU(2)$ gaugino masses for the solid (dark green) curve. We see that in
all cases the relic density depends very strongly on this parameter if it is
in the desired range (\ref{om2}), which is indicated by the shaded (light
green) band. In the first two cases the curve ends after dropping steeply, so
the desired relic density is again obtained close to the edge of the allowed
parameter space. In the last case the curve extends far beyond the point where
$\Omega_{\tilde \chi_1^0} h^2$ is in the right range. In fact, in this part of
the curve the relic density only depends relatively weakly on the plotted
parameter. However, in order to have this rather flat part of the curve
coincide with the desired range (\ref{om2}), one either needs a very heavy
LSP, well in excess of 1 TeV, which would lead to severe finetuning in the
Higgs sector; or one needs non--standard cosmology to raise the relic density
for parameters that are phenomenologically acceptable.

\begin{figure}[htb]
\begin{center}
\rotatebox{270}{\includegraphics*[width=13cm]{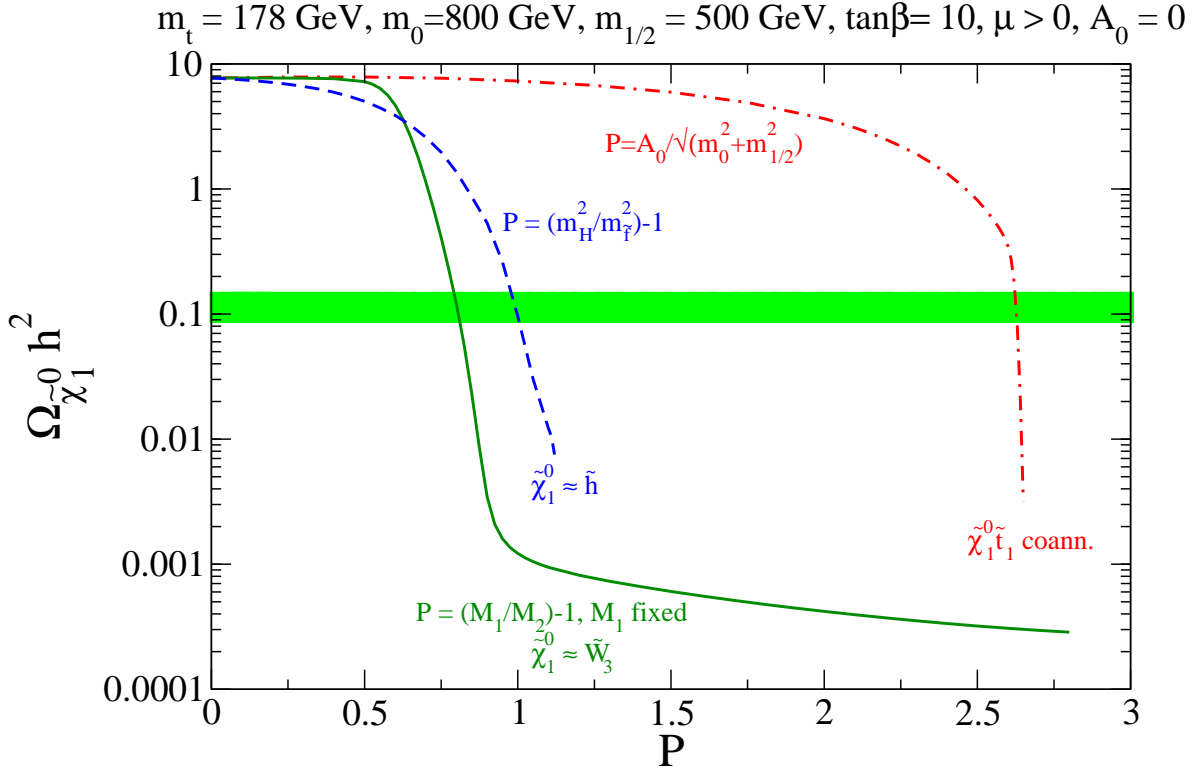}}
\caption{%
The $\tilde \chi_1^0$ relic density plotted as function of one parameter,
which measures the deviation from the indicated mSUGRA starting point. The
desired range (2) is indicated by the shaded (light green) band. See the text
for further details.}
\label{fig3}
\end{center}
\end{figure}

This brings me to the issue of non--standard cosmology. In the minimal
scenario one assumes that $\tilde \chi_1^0$ was in full thermal equilibrium
in the early Universe, which requires \cite{md24}
\begin{equation} \label{equil}
n_{\tilde \chi_1^0} \langle v \sigma_{\rm ann}^{\rm eff} \rangle > H = \sqrt{
  \frac {\rho_{\rm tot}} {3 M_{\rm Pl}^2} } \, .
\end{equation}
Here $n_{\tilde \chi_1^0}$ is the $\tilde \chi_1^0$ number density,
$\sigma_{\rm ann}^{\rm eff}$ is the effective sparticle to particle
annihilation cross section, which might include co--annihilation effects
\cite{md25}, $v$ is the relative velocity of the two annihilating sparticles,
$\langle \dots \rangle$ denotes thermal averaging, $H$ is the Hubble
parameter, $\rho_{\rm tot}$ is the total energy density of the Universe, and
$M_{\rm Pl} \simeq 2.4 \cdot 10^{18}$ GeV is the reduced Planck mass. Note
that both sides of this relation depend strongly on the temperature $T$. For
$T < m_{\tilde \chi_1^0}$ this dependence becomes exponential on the lhs,
whereas in standard cosmology it's only a power--law on the rhs. Clearly the
inequality can therefore not be satisfied at very low $T$.  The relic density
depends crucially at the ``freeze--out'' temperature $T_f$, where the
inequality becomes an equality, with lower $T_f$ corresponding to lower
$\Omega_{\tilde \chi_1^0}$ since $n_{\tilde \chi_1^0} \propto \exp(-m_{\tilde
  \chi_1^0}/T)$. ``Standard cosmology'' means that the expression for $H$
given in (\ref{equil}) holds, and that the Universe at $T \simeq T_f$ was
radiation--dominated with only SM degrees of freedom as relativistic
particles. This allows to compute the rhs in Eq.(\ref{equil}) as function of
$T$, and yields $T_f \simeq m_{\tilde \chi_1^0}/20$. Clearly we can increase
$T_f$, and hence $\Omega_{\tilde \chi_1^0}$, by increasing $H(T)$. Several
ways to do so have been suggested \cite{md26}, which could make a wino--like
$\tilde \chi_1^0$ with reasonable mass a good DM candidate.

Of course, a bino--like $\tilde \chi_1^0$ typically has too {\em large} a
relic density. Increasing $H(T)$ would make the situation even worse. In
principle it should also be possible to cook up scenarios where $H(T)$ is
reduced, but I am not aware of any studies along these lines. Another
possibility is to dilute the $\tilde \chi_1^0$ density after freeze--out by
releasing entropy from some late particle decay. The point is that one
actually calculates the {\em ratio} of $n_{\tilde \chi_1^0}$ and the entropy
density. In the standard calculation an absolute number for $\Omega_{\tilde
  \chi_1^0}$ is derived by assuming that the entropy density per comoving
volume remained constant for $T < T_f$. If a late decay significantly
increased the entropy density, it would reduce the $\tilde
\chi_1^0$--to--entropy ratio, and hence $\Omega_{\tilde \chi_1^0}$. In order
not to mess up Big Bang nucleosynthesis, this decay should happen at $T > 1$
MeV, but this still leaves several orders of magnitude in $T$, and twice as
many orders of magnitude in lifetime of the decaying particle, where this
mechanism could work. A good example for a late decaying particle is a hidden
sector field, nowadays called moduli. However, if this decaying particle
couples directly to $\tilde \chi_1^0$, these decays can {\em increase}
$\Omega_{\tilde \chi_1^0}$ \cite{md27}; recall that this is desirable if
$\tilde \chi_1^0$ is higgsino-- or wino--like.

Finally, I should mention that there are viable SUSY DM candidates other than
the lightest neutralino. Every SUSY model must contain a gravitino. There are
at least three different sources of gravitinos in the early Universe: direct
production from the thermal plasma; production from the decay of MSSM
sparticles before the latter froze out; and decay of the lightest
visible--sector sparticle at $T < T_f$. It is therefore not surprising that
one can arrange things such that one gets the right relic density for pretty
much any combination of visible--sector soft breaking parameters, e.g. by
choosing appropriate values of the gravitino mass and of the reheat
temperature after inflation. Since gravitino DM is treated in three
contributions to these Proceedings \cite{md28}, I will not discuss it any
further, except for making the obvious remark that it is experimentally
impossible to prove that gravitino DM indeed exists; its couplings to ordinary
matter are just too weak.\footnote{One {\em can} perhaps {\em dis}prove this
  possibility in certain cases, by studying gravitinos in the lab \cite{md28}.
  However, a failure to disprove does not make a proof.} This is true also for
another SUSY DM candidate, the axino \cite{md29}.

\section{Claimed WIMP detections}

This brings me to the issue of WIMP detection. In fact, there are several
particle physics observations that have been interpreted as positive evidence
for particle DM. 

The first is the DAMA observation of a statistically significant annual
modulation in the observed event rate, interpreted as being due to the
scattering of ambient DM particles off the nuclei in the detector
\cite{md30}. More recent stringent limits from other direct WIMP searches
\cite{md7} prove that the DAMA signal cannot be due to a SUSY WIMP. As far as
I know, some (even) more exotic DM particles might still be compatible with
all data \cite{md32}, but I personally am very sceptical about this
observation.

The other positive evidence all comes from the observation of fluxes of
energetic particles in the vicinity of Earth. In particular, it has recently
been pointed out \cite{md33} that an excess in photons with energy in the few
GeV range can be explained by WIMP annihilation into $b \bar b$ pairs; this
can even be described by mSUGRA, if one allows a ``boost factor'' in the
annihilation rate, which could e.g. be due to small--scale clumpiness of the
halo. This scenario also improves the description of the positron flux at a
few tens of GeV, and of the antiproton flux below 1 GeV. Moreover, the photon
data are good enough to reconstruct the DM distribution in our galaxy; one
finds structures (rings) at radii which coincides with structures in the
visible galaxy (a gas ring and a ring of stars). This sounds quite impressive.
However, I find it difficult to assess the significance of this observation.
The crucial question is how reliable the ``SM'' predictions for the background
fluxes are. Among other things, they rest on the assumption that the CR fluxes
(of protons and electrons) are essentially the same everywhere in our galaxy.
We know that the distribution of hot gas, of magnetic fields, of starburst
regions etc is quite inhomogeneous; given that the excess in all cases is only
a factor of a few, I would personally not want to bet money that the observed
deviations are indeed due to WIMPs.

Finally, an excess of 511 keV photons from near (but not right at) the
galactic center has been interpreted as light DM particles $\chi$ annihilating
or decaying into $e^-e^+$ pairs \cite{md34}. If $m_\chi \leq 100$ MeV, the
positrons would be slowed down sufficiently fast to annihilate (mostly) at
rest; this bound has very recently been lowered to 20 MeV by considering
emission of photons during the decay or annihilation \cite{beacom}.  $\chi$
could not be the lightest neutralino in an R--parity conserving MSSM (with
non--universal gaugino masses, to allow such a light neutralino), since its
annihilation cross section would be much too small. A decaying particle does
not fit the angular distribution of the signal very well \cite{md35}.  On the
other hand, there might be some connection with extended, $N=2$, SUSY
\cite{md36}.  However, astrophysical explanations of this observation have
also been suggested \cite{md37}.

\section{Summary and conclusions}

The lightest neutralino $\tilde \chi_1^0$ remains the best motivated DM
candidate. It remains viable even in the simplest models of both particle
physics (mSUGRA) and cosmology (standard cosmology up to temperatures of a few
dozen GeV at least). Imposing the relic density constraint in this framework
greatly reduces the size of the allowed parameter space, but does {\rm not}
reduce most lower bounds on sparticle and Higgs boson masses significantly.
Scenarios with little finetuning in both the Higgs and DM sector are still
allowed, if the top mass is near the upper end of its experimentally
determined range and the lightest Higgs boson is close to the lower bound on
its mass, as indicated by the (not very compelling) ALEPH hint of a Higgs
discovery \cite{md38}.

Non--universal SUGRA models do not increase the fraction of the (enlarged)
parameter space that satisfies the relic density constraint, nor do they
reduce the finetuning required to satisfy this constraint unless one also
modifies cosmology (in the direction of enhancing the relic density). Other
SUSY DM candidates exist. The gravitino is as well motivated from the particle
physics side as the neutralino, but unlike the case of thermal WIMPs, nothing
singles out the desired relic density even on a logarithmic scale.

Several distinct experimental observations have been interpreted as positive
evidence for (mutually incompatible) WIMP Dark Matter, but only one them
\cite{md33} can be interpreted as evidence for a neutralino WIMP without
violating other constraints. This particular observation looks quite
intriguing, but I wouldn't call it compelling. 

This raises the question what kind of experimental evidence would be required
to make a compelling case for neutralino Dark Matter. I believe that
observation of a WIMP signal by itself will not be sufficient. Astroparticle
physics experiments cannot discover supersymmetry before the LHC does,
contrary to claims one sometimes sees in the literature. To be sure, if they
(and we) are lucky, a compelling WIMP signal might be found before he LHC
commences operations, but it will be impossible to convince people that this
WIMP is a superparticle; other candidates exist already now, and many more are
sure to be invented as soon as a signal is established. Collider experiments
will be necessary to establish that this WIMP is indeed a superparticle. Such
experiments should eventually also be able to determine the masses and
couplings one needs to know in order to compute the total LSP annihilation
cross section \cite{md39}, which will allow to calculate its relic density in
a variety of cosmological models. The combination of these results from
colliders with WIMP detection experiments will thus allow us to probe the
Universe at a temperature some four or five orders of magnitude above that at
the onset of nucleosynthesis, which currently is the earliest well established
epoch.  Collider physics and astroparticle physics should therefore not be
seen as competitors, but as partners.

\section{Acknowledgements}
I thank: my collaborators Abdelhak Djouadi, Jean--Loic Kneur and Pietro Slavich
for the cooperation in those results presented here that are new; Xerxes Tata
and Jounghun Lee for useful conversations; and the KIAS school of physics for
hospitality. 

\bibliographystyle{plain}

\end{document}